\documentclass[pra,aps,twocolumn,superscriptaddress,10pt,floatfix]{revtex4}

\usepackage{amsmath}
\usepackage{amssymb}
\usepackage{graphicx}
\usepackage{microtype}
\usepackage{subfigure}
\usepackage[svgnames]{xcolor}
\usepackage[pdftex]{hyperref}
\hypersetup{colorlinks=true,linkcolor=blue,citecolor=ForestGreen,urlcolor=DarkOrchid}
\graphicspath{{images/}}

\newcommand{\lsu}{\affiliation{Department of Physics \& Astronomy, Louisiana State University, Baton Rouge, LA, USA}}
\newcommand{\utmb}{\affiliation{Department of Preventive Medicine \& Population Health, University of Texas Medical Branch, Galveston, TX, USA}}
\newcommand{\ubc}{\affiliation{Outer Space Institute, University of British Columbia, Vancouver, CA}}
\newcommand{\spacex}{\affiliation{Aerospace Medicine and Vestibular Research Laboratory, The Mayo Clinic, Arizona; Scottsdale, AZ, USA;}}
\newcommand{\lsuhsc}{\affiliation{Department of Internal Medicine, Louisiana State University Health Science Center, Baton Rouge, LA, USA}}
\newcommand{\nasa}{\affiliation{Astronaut Office, NASA Johnson Space Center, Houston, TX, USA}}

\begin{document}
\title{Tetrahedral Human Phantoms for Space Radiation Dose Assessment}

\author{Megan E. Chesal}
\lsu
\author{Rebecca S. Blue}
\spacex
\utmb
\author{Serena M. Aunon-Chancellor}
\utmb
\lsuhsc
\nasa
\author{Jeffery C. Chancellor}
\lsu
\utmb
\ubc
\email{jchancellor1@lsu.edu}

\begin{abstract}
Space radiation remains one of the primary hazards to spaceflight crews. The unique nature of the intravehicular radiation spectrum makes prediction of biological outcomes difficult. Recent advancements in both Monte Carlo simulations and computational human phantom developments have allowed for complex radiation simulations and dosimetric calculations to be performed for numerous applications. In this work, advanced tetrahedral-type human phantoms were exposed to a simulated spectrum of particles equivalent to a single-day exposure in the International Space Station in Low Earth Orbit using 3D Monte Carlo techniques. Organ absorbed dose, average energy deposition, and the whole-body integral dose was determined for a male and female phantom. Results were then extrapolated for three long-term scenarios: a 6-9 month mission on the International Space Station, a 2 year mission on the International Space Station, and a 3-year mission to Mars. The whole-body integral dose for the male and female models were found to be 0.2985 $\pm 0.0002$ mGy/day 0.3050 $\pm 0.0002$ mGy/day, respectively, which is within 10\% of recorded dose values from the International Space Station. This work presents an approach to assess absorbed dose from space-like radiation fields using high-fidelity tetrahedral-type computational phantoms, highlighting the utility of complex models for space radiation research. 
\end{abstract}

\maketitle
\section*{Introduction}

The potential risk to human health incurred from prolonged exposure to the space radiation environment remains a primary limiting factor for long-duration space missions \cite{chancellor2014space,chancellor2018limitations}. The space radiation environment is a complex combination of fast-moving ions derived from all atomic species found in the periodic table \cite{chancellor2018limitations}. These ionized nuclei have sufficient energy to penetrate a spacecraft structure and cause deleterious biological damage to astronaut crews and other biological material, such as cell and tissue cultures \cite{chancellor2018limitations, chancellor2021everything}. To date, our understanding of the nature of this risk, and impact to human health, remains incomplete. Further, there continues to be disparity between radiobiological studies and actual health outcomes seen in astronaut crews \cite{chancellor2018limitations, kennedy2014biological}.

Current space radiobiology studies primarily rely on animal model surrogates, translating interspecies findings in an attempt to predict clinical outcomes for human physiology \cite{kennedy2014biological}. These animal studies have contributed to our understanding of disease mechanisms; however, their relevance to extrapolating outcomes in humans remains uncertain \cite{williams2010animal,nelson2016space,hackam2006translation,perel2007comparison,chancellor2018limitations}. It is believed that sufficient exposure to the complex space radiation environment could cause multi-organ tissue damage, e.g. inhibiting cell regrowth and repair mechanisms. Recent studies have demonstrated that the biological response and disease pathogenesis seen after exposure to space radiation is unique to a nonhomogeneous, multi-energetic dose distribution \cite{chancellor2018limitations, chancellor2021everything}. Thus, high-fidelity simulation of the complexity of the space environment and the human body is critical for accuracy in radiobiological experimental outcomes and terrestrial research. Historical calculations of space radiation biological impact have relied on more simplistic representations of human anatomy, the environment, and space radiation penetration and deposition in the body or individual organ systems due to previous limitations of calculations, phantom fidelity, or prediction of intravehicular spectra and dosage \cite{Reitz2009, Badhwar2002, Yasuda2009, chancellor2017targeted}. This has limited the success of dose prediction or in correlation to clinical outcome. Advancements over the past few decades have provided significant improvements in modeling and computational capabilities; however, many calculations are still performed using simplistic simulations or models, which can hinder accurate dose calculations \cite{slaba2010utilization}. 

In this work, we present an approach involving computational tetrahedral-type human phantoms and their use in high-resolution and accurate dosimetry calculations for simulated exposure to the intravehicular space radiation environment similar to that incurred by astronauts onboard the International Space Station (ISS). Our goal is to numerically predict the dose in major organ and tissue volumes after a whole-body exposure to the space radiation environment. Use of high-fidelity human phantoms for more accurate computational reproduction of human space radiation exposures may help elucidate how the relative distribution of energy deposition and subsequent stress on organ systems such as the brain, lungs, and vasculature may contribute to overall dose toxicity and potential health outcomes related to space radiation exposure. 

\begin{figure}[ht]
\includegraphics[width=60mm,keepaspectratio]{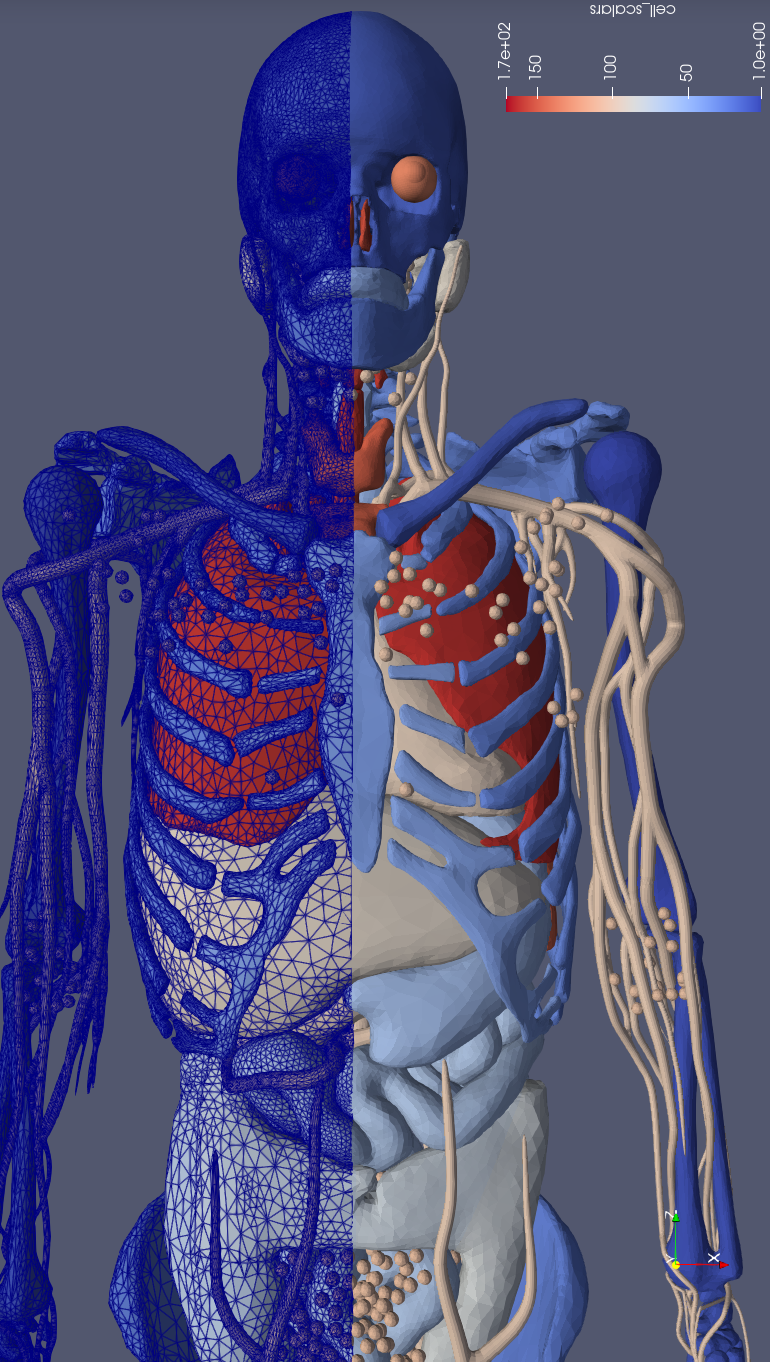}
\caption{{\bf 90th Percentile Female Tetrahedral-Type Phantom.} Tetrahedral-type phantoms are a progression of surface-type phantoms, with surface meshes replaced by volume tetrahedrons. Tetrahedral-type geometry is easily incorporated in Monte Carlo codes and leads to faster radiation transport calculations while still preserving the original complexity of surface phantoms. Shown here is the female tetrahedral-type phantom used in our calculations. In this view, the skin, muscles, and residual tissues are hidden to showcase internal anatomy structures. The right side of the phantom is visualized with smooth surfaces, while the left side is visualized as a surface with edges, showing the tetrahedrons that comprise the different model regions.}
\label{fig:tetragonal_female}
\end{figure}

\section*{Background}
The space radiation environment is primarily comprised of a complex mixture of energetic radiation particulates that originate from both the sun and from outside of our solar system. Radiation emitted from the sun is comprised of electromagnetic radiation as well as the solar wind, which consists of protons and alpha particles and a small percentage of heavy-charged particles ($Z>2$). Galactic Cosmic Rays (GCR) originate from outside of our solar system and are believed to be the remnants of supernova events \cite{simpson1983elemental}. Additionally, the low Earth orbit (LEO) radiation environment includes the Van Allen belts, toroidal regions comprised of electrons and protons. Vehicles (and crew) passing through these belts will incur additional exposure.

In nominal conditions within a shielded vehicle, the primary source of radiation of concern to  astronauts during long-duration deep space spaceflight is GCR \cite{chancellor2018limitations}. The GCR spectrum is comprised mainly of protons and helium ions, though a smaller component, approximately one percent, consists of heavier and highly energetic nuclei with charges equal to or greater than atomic number $Z=3$. While any ion species can be found in the GCR spectrum, the most prevalent ion species encountered are at or below Z = 28 \cite{chancellor2018limitations,simpson1983elemental}. Although the contribution of these heavier ions to the total spectrum relative to protons and helium ions may be some orders of magnitude smaller, their potential biological impact is of equal or greater value \cite{walker2013heavy,cucinotta2006evaluating}. Energies of particles in the GCR spectrum can vary from kilo-electronvolts (keV) to several tera-electronvolts (TeV), energies sufficient to enable GCR particles to penetrate spacecraft shielding and interact with crewmembers \cite{chancellor2018limitations,chancellor2017targeted}. As GCR nuclei travel through spacecraft shielding, interactions will occur with the shielding medium. Such interactions can instigate energy loss mechanisms; in some cases, heavy ions may fragment into lighter ions with greater potential for biological damage or destruction than the original heavy ion. This creates a unique intravehicular (IVA) spectrum inside the spacecraft. The LEO IVA spectrum can also include trapped charged particles due to the magnetic field surrounding the Earth; the relative contribution of these particles depends on the orbital flight path and how long a given vehicle remains in a region of trapped particles. For example, in certain trajectories the Van Allen belts can contribute significantly to total absorbed dose. While more shielding could be added to a vehicle to reduce the fluence of particles, payload lift capabilities limit the amount of material available \cite{chancellor2018limitations}. Additionally, studies have shown that dramatically increasing the shielding provides only limited reduction in particle fluence and could, in fact, increase the IVA radiation environment complexity  \cite{chancellor2021everything}.



The biological impact of radiation depends on a variety of physical and biological factors, including absorbed dose, dose rate, exposure time, particles species, material, and genetic variability. Computational simulations provide the capability for dosimetric calculations to be performed for simple and complex radiation spectra, either as a supplement to or in place of a physical experiment. Dose deposition in biological models can be simulated; however, to fully represent the complexity of such systems and environments, high resolution models, or phantoms, should be utilized for a more accurate description of dose topology and unique astronaut morphology \cite{bahadori2011effect, kim2020icrp}. Voxel phantoms, comprised of volume pixels, have long been used to assess absorbed dose for various radiation exposures, both terrestrially and extra-terrestrially \cite{kramer2006max06}. Voxel phantoms, however, have stair-stepped surfaces rather than continuous surfaces which has been shown to result in inaccurate dose calculations \cite{hickson2014effect}. Additionally voxel phantoms are rigid and cannot be deformed, limiting their utility to be used in more complex scenarios \cite{lee2007hybrid}. Advancements in computational phantoms have allowed for the development of more complex models, allowing higher fidelity recreation of fine structures. The current premier phantoms are surface-type phantoms, where delineated organs and regions are defined by surface meshes \cite{lee2007hybrid, lee2007nurbs, na2010deformable}. Work has been done investigating space radiation using surface-type phantoms that utilize non-uniform rational B-spline (NURBS) modeling with a variety of computational software \cite{bahadori2012dosimetric,bahadori2013comparative,bahadori2011effect}. However, NURBS-based models cannot be directly implemented into 3D Monte Carlo software, requiring a revoxelization of the model before simulation and losing some of the advantages of NURBS modeling in the process \cite{lee2007nurbs}.

\begin{figure}[ht]
\includegraphics[width=\linewidth,keepaspectratio]{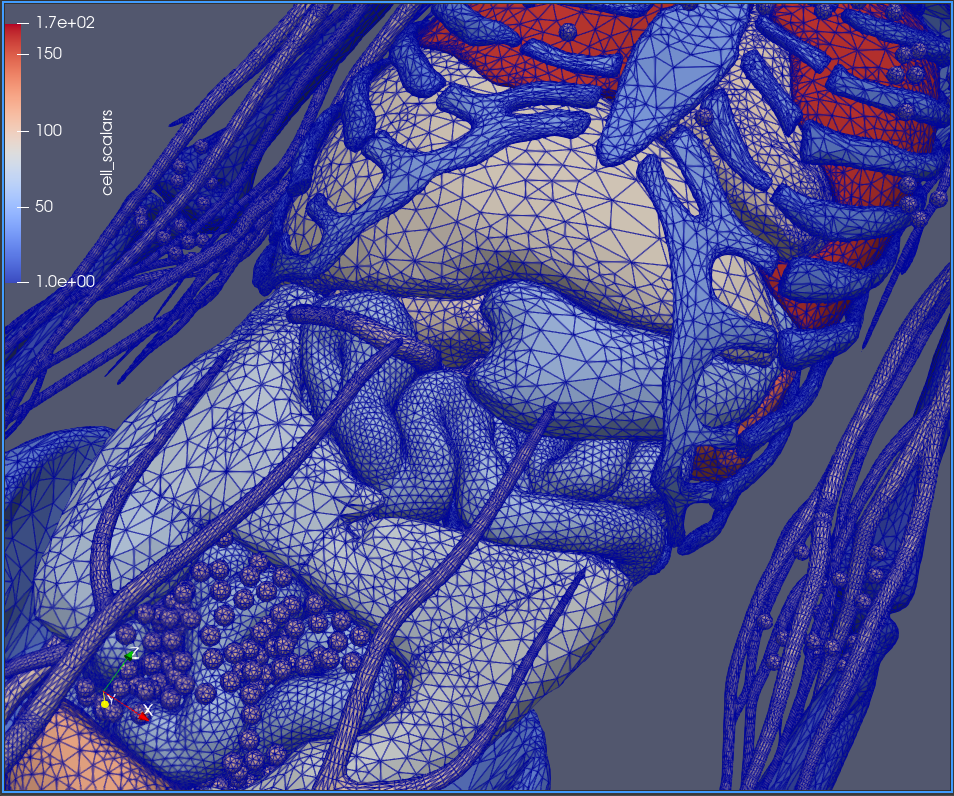}
\caption{{\bf Tetrahedral-Type Phantom Midsection Detail} Tetrahedral-type phantoms are constructed with volumes filled with varying sized tetrahedrons. In the image, the lines on the organs represent the various tetrahedrons used to define that volume. Volumes and regions defined in this manner depend only on the size of the tetrahedrons, which is unlimited by modeling technique. While in large organs finely detailed tetrahedrons may not be needed, small or complex structures, such as mucosal linings, benefit greatly from a smaller-volume tetrahedron approach and can be modeled at the micrometer level.}
\label{fig:half-and-half}
\end{figure}

In 2020, the International Commission on Radiological Protection (ICRP) published Report 145 which introduced two phantoms, an adult male and female polygonal surface mesh-reference computational phantom (MRCP), for use in dosimetric calculations \cite{kim2020icrp}. Utilizing the IRCP MRCP models, Lee et. al. developed tetrahedral-type phantoms representing the 90th percentile height and weight for male and female individuals for the worldwide Caucasian population \cite{lee2019percentile}. Tetrahedral-type phantoms are advantageous for radiation transport simulations; they can be developed from medical images and detailed physical structures, such as lymph nodes, can be incorporated into the final phantom (as shown in  Figure \ref{fig:tetragonal_female}). Deformation techniques, used to adjust select body regions for greater fidelity or to change a phantom positioning, can additionally be applied \cite{kim2011polygon,yeom2014tetrahedral}. Tetrahedral phantoms are created by applying mathematical techniques, such as Delaunay triangulation, to a surface-mesh that convert phantom structures to volumes defined by a series of tetrahedrons \cite{si2009quality}. Figure \ref{fig:half-and-half} demonstrates how complex and minuscule surfaces can be defined with smooth edges; boundaries for organs are well defined in this manner. Further, these high-fidelity models are amenable to the incorporation of three-dimensional Monte Carlo particle transport software and have computational times comparable to voxel-based phantoms \cite{yeom2014tetrahedral}.
Here we utilize the tetrahedral-type phantoms developed by Lee et. al. and incorporate the Particle and Heavy Ion Transport Code System (PHITS), a three-dimensional radiation transport Monte Carlo platform with an event generator mode capable of simulating particles up to 200 GeV/n \cite{sato2018features,boudard2013new,hirayama2005egs5}. Incorporation of  the advanced phantoms into PHITS software can allow for high-fidelity computation of dose deposition during spaceflight.

\section*{Methods}
\begin{figure*}[ht]
\includegraphics[width=150mm,keepaspectratio]{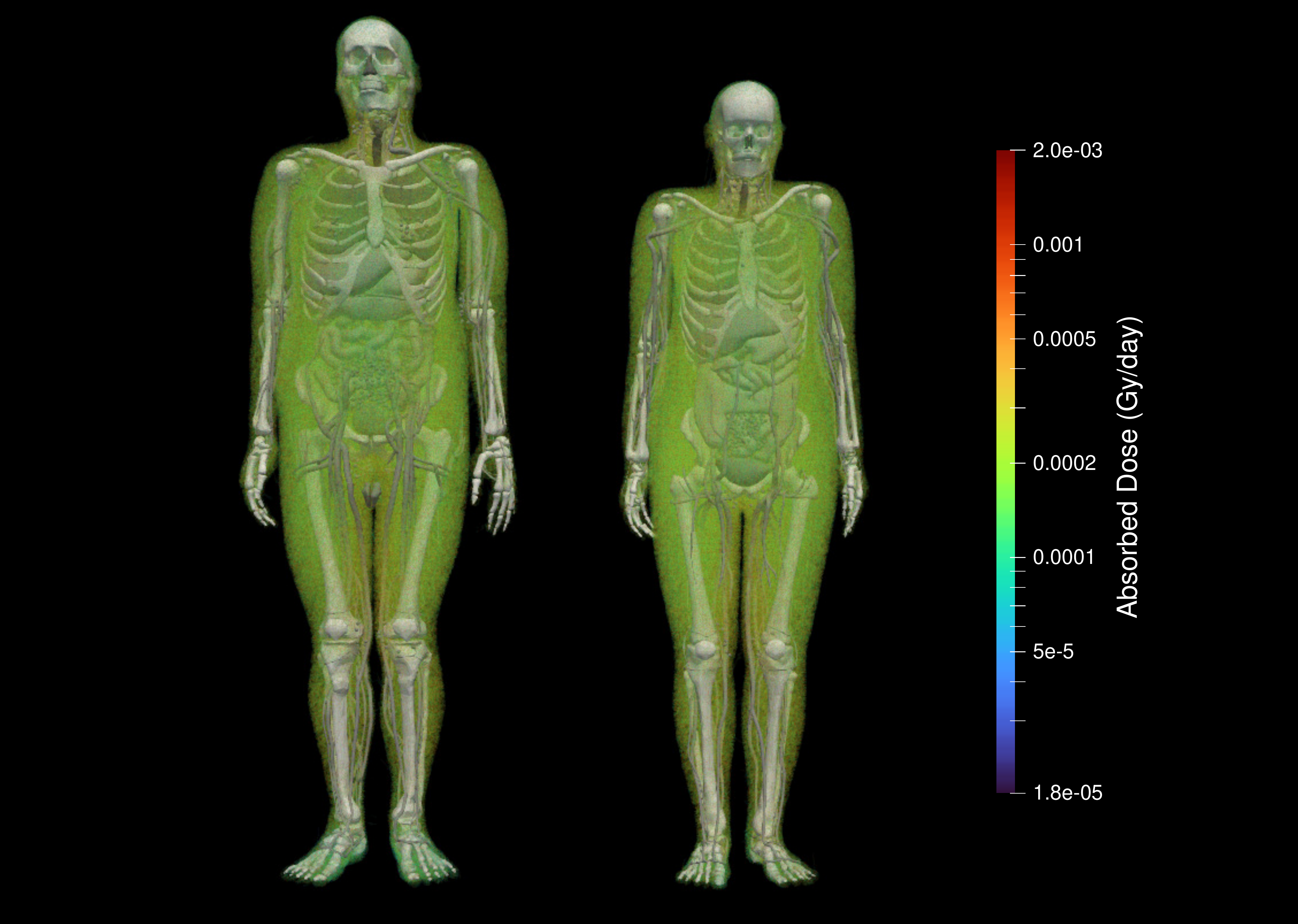}
\caption{{\bf Male and female phantoms.} The 90th percentile mass and height male (left) and female (right) phantoms were exposed to a simulated IVA spectrum representing space radiation dosage of roughly one full day of exposure in space. Phantoms are shown without skin, muscle, and soft tissue to show internal organs and skeletal frame. The three-dimensional dose distribution of a fraction of the overall dose is overlaid on the phantoms to show the anatomical relation of the dose distribution. All region doses in both phantoms are on the order of $\mu$Gy.}
\label{fig:mf-full}
\end{figure*}

Male and female 90th percentile mass and height surface phantoms were tetrahedralized by defined regions. Each phantom was individually exposed to a simulated IVA space radiation spectrum. To recreate the IVA spectrum, each individual simulated particle was transported through the phantom as an independent event, with energy reflecting IVA linear energy transport (LET) and particle spectrum as measured onboard the ISS Columbus Module in a "moderator block" technique previously described by Chancellor et. al. \cite{chancellor2017targeted}. The spectrum generation method proposed by Chancellor et. al. uses charged particle interactions, specifically energy loss and spallation, to change the output of a heavy ion beamline to a more complex spectra. In the design proposed, 1 GeV iron ions pass through a moderation material of variable thickness, fragmenting the ions and reducing the energy of the primary ions and the produced ions such that the LET spectra and charged particle distribution produced matches measurements taken on ISS. The results were compared to measurements taken onboard ISS and found to be within excellent agreement \cite{chancellor2017targeted, kroupa2015semiconductor}. The moderation device was modeled in the 3D Monte Carlo software PHITS and the resulting spectra was recorded a meter behind the end of the block. The particle positions and directions were then reassigned onto a series of planes surrounding the phantom to create a homogeneous exposure representative of the average environment on ISS. 

All simulations were performed via PHITS and utilized high performance computers with parallel computational capacity. The average absorbed dose in each defined region was determined and additional tallies were set to record the three-dimensional dose distribution at two resolutions, 0.8~mm$^3$ and 0.2~cm$^3$. 

Absorbed dose and average energy deposition was determined for all organ sites. Absorbed dose was calculated with the standard definition of energy deposition per unit mass for defined volume, and the whole body dose was calculated using the integral dose, in which the total energy deposited over the phantom was divided by the total mass of the phantoms \cite{icrp1990recommendations}. Due to software limitations at the time of computation, the material surrounding the phantoms was approximated by low density air, with a density of 10$^{-6}$ g/cm$^3$, to minimize energy loss of the particles before interaction with the phantoms.

Figure \ref{fig:mf-full} represents the 0.2 cm$^3$ measured dose distribution, where dose external to the phantom is disregarded. Average regional doses were exported with relative error for analysis. Average organ doses were determined with a weighted average of the recorded organ components, where weighting utilized the ratio of the volume contribution of a region to the total organ volume. Specific organ groups, such as the skeletal system and the lymphatic nodes, were calculated as the average contribution of each component for a single total group value. Organ system doses were calculated as the sum of their organ components, and the total body dose was then calculated as the sum of the organ system and extraneous organs not previously included in a system. The average energy deposited for these same categories was determined by using the mass of each region. 

Finally, daily dose exposures were used to calculate predictive doses for a 6-9 month, a 2-year, and a 3-year mission for male and female crewmembers. A 6-9 month time period would be typical for exposure times during current missions on the ISS while a 2-year mission would be an extended orbital mission beyond current historical experience. A 3-year mission is the anticipated time frame for an interplanetary exploration mission to Mars. For a 3-year mission to Mars, the dose can be approximately determined by assuming an average Earth to Mars transit time of 200 days, 500 days of exploration on the Mars surface, and 200 days for the return to Earth \cite{drake2010human}. GCR fluence is known to be 2-3 times greater outside of LEO, so a factor of 2.5 was used to increase the particle fluence. Prior literature by Zeitlin et. al. describes the interplanetary radiation spectrum as roughly equivalent to that of LEO; as such, calculations utilized dose rates as described by Zeitlin et al. using the fluences described above \cite{zeitlin2019comparisons}. The dose incurred from a Mars expedition was then calculated as the sum of the absorbed dose from each stage of the mission.  Calculated doses are based on ISS reference IVA radiation in LEO and thus are not fully reflective of a potential interplanetary exposure; despite this limitation, cumulative doses were calculated and are presented below for further discussion.

\section*{Results}
The reference computational phantoms were organized into 170 individual regions. Regions ranged from individual organs or structures (e.g. brain, spinal cord, thyroid) to specified sections of an organ or structure (e.g. glandular and adipose breast tissue, spongiosa in the proximal femur). For each of these individual regions, the average absorbed dose and energy deposited are presented in Tables ~\ref{tablemale} and ~\ref{tablefemale} for male and female phantoms, respectively. 

The single day whole body absorbed dose for the male and female computational phantoms were determined to be 0.2985~$\pm 0.0002$ mGy/day and 0.3050~$\pm 0.0002$ mGy/day, respectively, as demonstrated in Figure~\ref{fig:mf-full}. Prior studies measuring IVA dose deposition in the ISS Columbus module have reported dose values of 0.286 mGy/day\cite{berger2017dosis}; thus, calculated whole body doses are within 6.6\% of actual measured ISS IVA values. A few systems of interest will be discussed below. 

\subsubsection*{Gastrointestinal System}
The complexity of the gastrointestinal system required that 49 individual regions be mapped out for both male and female computational phantoms. For organ totals, the relevant regions were averaged as described above. Regions in the oral cavity were specifically defined including the oral mucosa on the tongue and cheeks. These regions were averaged by their mass component and further averaged with extrathoracic regions representing facial structures for a general oral and facial cavity subset. The colon was further divided into the ascending colon, the right transverse colon, the left transverse colon, the descending colon, the sigmoid colon, and the rectum wall. Those regions were averaged for a total mean colon dose.  

\begin{figure*}[ht]
\includegraphics[width=150mm,keepaspectratio]{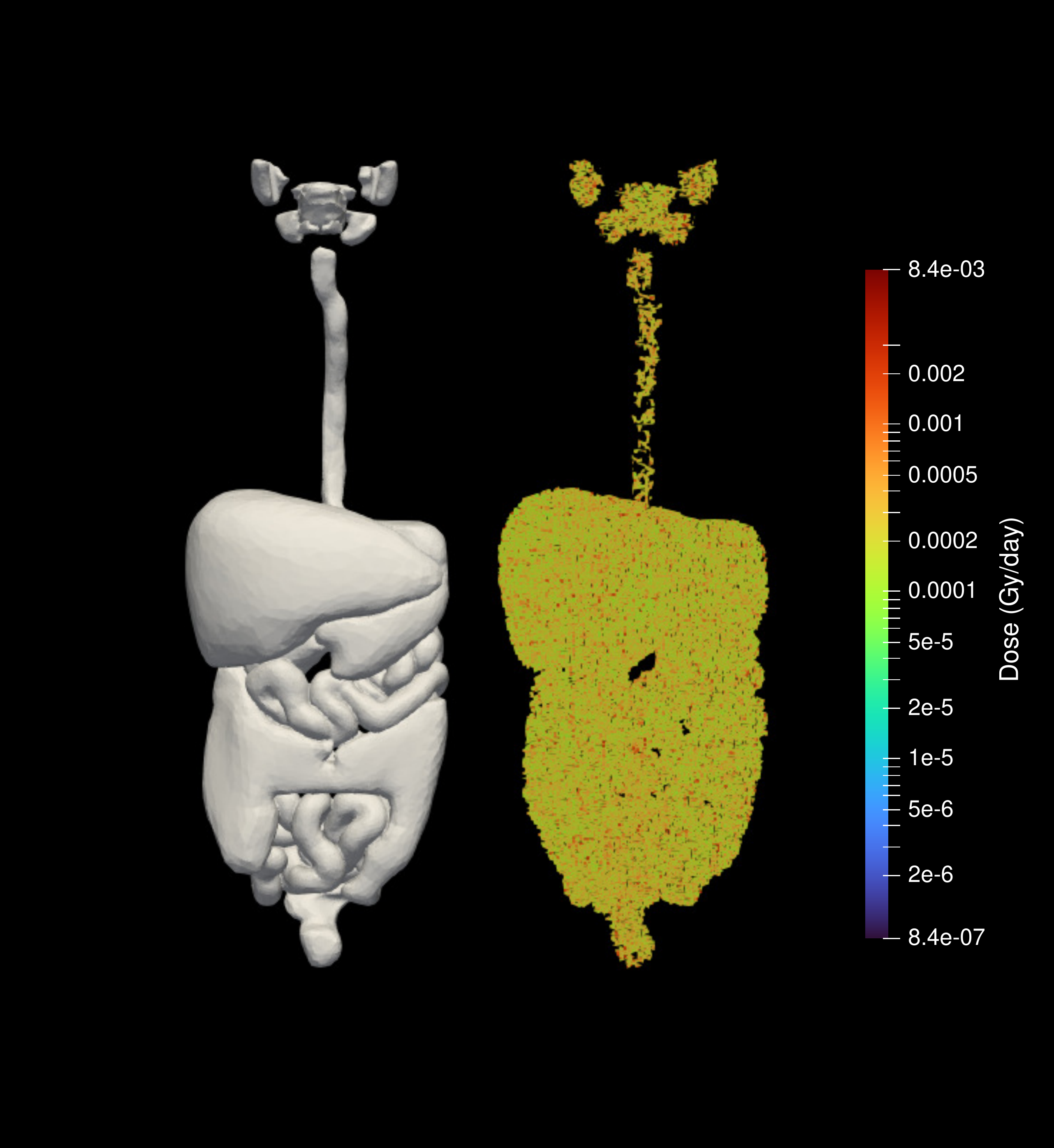}
\caption{{\bf Gastrointestinal System of the Female Phantom.} The gastrointestinal system model is comprised of 49 regions, shown on the left. The corresponding approximate absorbed dose distribution from the simulated LEO exposure is shown on the right. Cooler colors represent a lower absorbed dose while warmer colors represent to a higher dose.}
\label{fig:f-gastro}
\end{figure*}

The total gastrointestinal system dose was determined as the integral sum of the average doses to the following regions: the oral cavity, esophagus, stomach, small intestines, colon, salivary glands, tonsils, liver, gallbladder, and pancreas. Both male and female phantoms demonstrated highest gastrointestinal organ dose distribution in the salivary glands and the tonsils. For the female computational phantom, the gastrointestinal system received a total dose of 0.2842 $\pm{\:0.0004}$ mGy/day, while the male computational phantom received a total gastrointestinal dose of 0.2705 $\pm{\:0.0003}$ mGy/day. The spatial absorbed distribution of the female phantom can be seen in Figure \ref{fig:f-gastro}. 

\subsubsection*{Circulatory System} 
The heart wall and blood contents of the heart were delineated into two separate regions. All of the arteries and veins mapped out in the computational phantoms are grouped into a singular artery or vein classification, with no distinction for the different vessels or location in the body. For blood vessel mean doses, arterial and venous doses were averaged together. The total circulatory system mean doses were calculated as the sum of the mass average of the heart and the average of the major vessels. 

\begin{figure*}[ht]
\includegraphics[width=150mm,keepaspectratio]{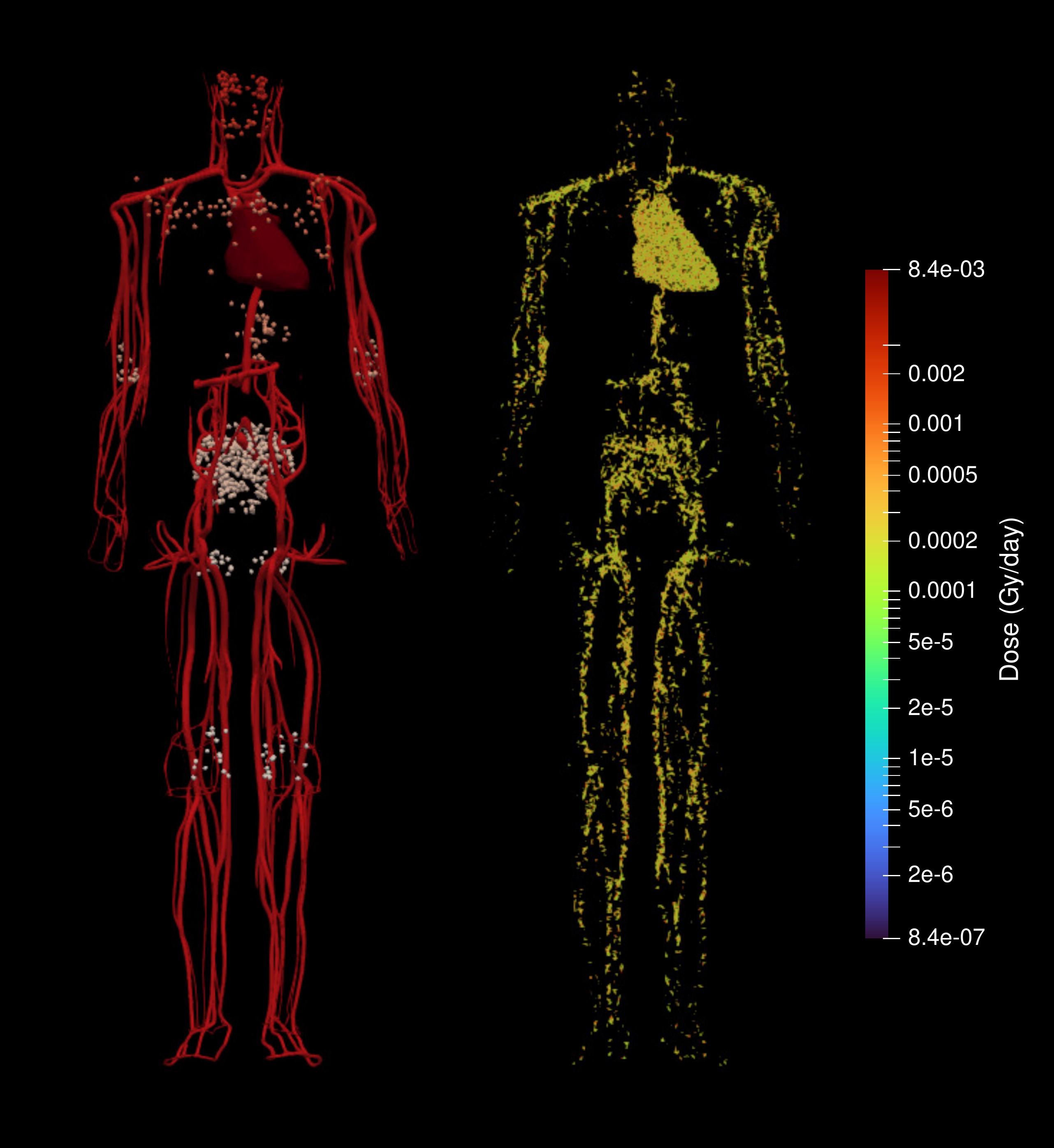}
\caption{{\bf Circulatory and Lymphatic Systems of the Female Phantom.} Shown on the left is the circulatory and lymphatic systems of the female phantom model and the corresponding absorbed dose distribution from simulated LEO exposure is on the right. Larger energy deposition tracks are represented by warmer colors on the dose distribution figure.}
\label{fig:f-circulatory}
\end{figure*}

For the female computational phantom exposure, the heart received an average dose of 0.3017 $\pm{\:0.0009}$ mGy/day, averaged over the heart wall and the heart contents.  The arteries and veins received an average dose of 0.3167 $\pm{\:0.0006}$ mGy/day. For the male computational phantom exposure, the heart received an average dose of 0.2987 $\pm{\:0.0009}$ mGy/day, averaged over the heart wall and the heart contents.  The arteries and veins received an average dose of 0.3084 $\pm{\:0.0005}$ mGy/day. In total, the female phantom circulatory system received a dose of 0.3107 $\pm{\:0.0005}$ mGy/day, and the male phantom circulatory system received a dose of 0.3041 $\pm{\:0.0004}$ mGy/day. The spatial absorbed dose distribution for female phantom circulatory and lymphatic systems for this LEO exposure are shown in Figure \ref{fig:f-circulatory}.

\subsubsection*{Brain}
\begin{figure*}[ht]
\includegraphics[width=150mm,keepaspectratio]{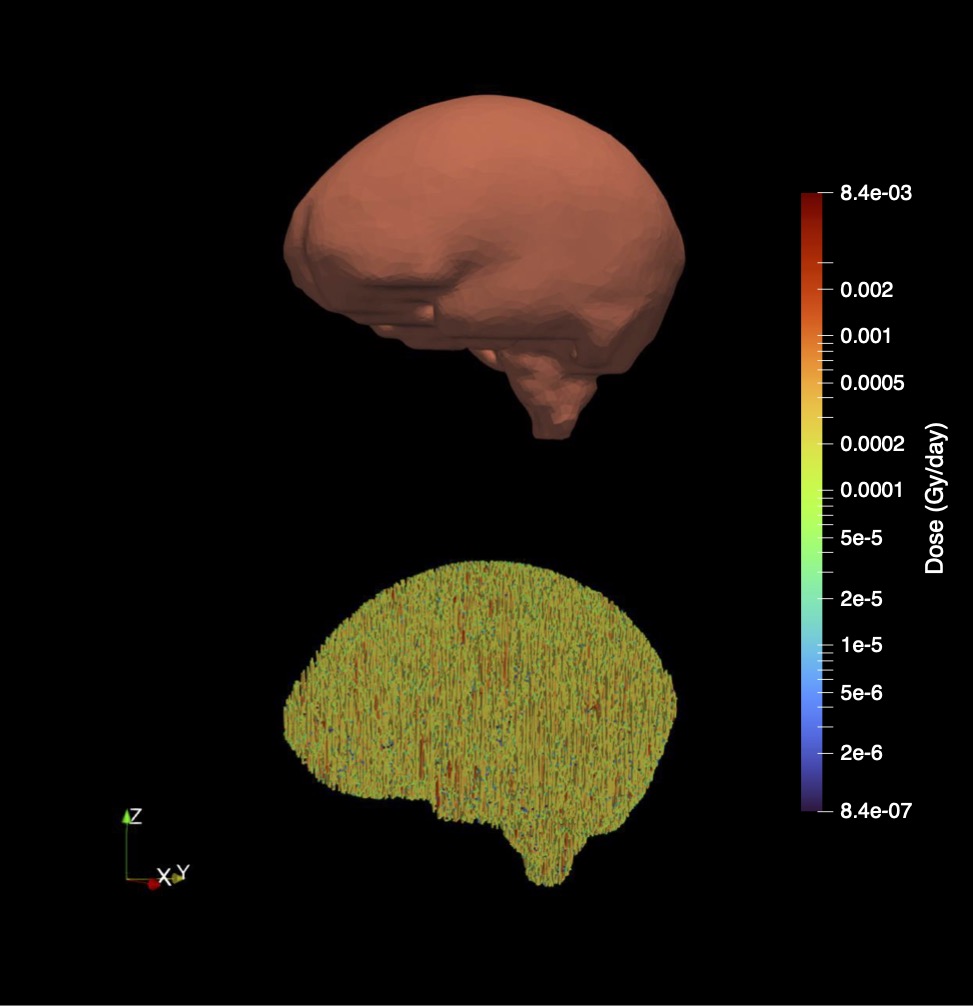}
\caption{{\bf Brain of the Female Phantom.} Above is the brain from the female model while below is the absorbed dose distribution from the simulated LEO exposure. Large energy deposition tracks from heavy charged particles can be seen in the image, where warmer colors represent larger doses.}
\label{fig:f-brain}
\end{figure*}

For the female computational phantom, the average dose to the brain (excluding eyes, pituitary gland, and spinal cord) was 0.2899 $\pm{\:0.0009}$ mGy/day; for the male computational phantom, the average dose was 0.259 $\pm{\:0.001}$ mGy/day. The brain was not further delineated into individual components; thus, average doses to particular intracranial subregions (for example, individual lobes) is unknown. The absorbed dose distribution from this exposure for the female phantom brain can be seen in Figure \ref{fig:f-brain}. 

\subsubsection*{Cumulative Dose Predictions for Long-Duration Spaceflight}

In a 6-9 month exposure for a female astronaut, predicted cumulative doses were calculated to be $51.2-76.8$ mGy for the gastrointestinal system, $55.9-83.9$ mGy for the circulatory system, and $51.9-77.9$ mGy to the brain. For a male astronaut, the predicted total doses for the same organ sites are $48.7-73.0$ mGy, $54.7-82.1$ mGy, and $46.5-69.9$ mGy, respectively. For whole body exposure, the predicted dose for both female and male astronauts is $54.9-82.4$ mGy and $53.7-80.1$ mGy, respectively. 

For an extended 2-year mission on ISS, the predicted cumulative doses for the female phantom were $204$ mGy for the gastrointestinal system, $223$ mGy for the circulatory system, and $207$ mGy to the brain. For a male astronaut, the predicted doses for the same organ sites are $196$ mGy, $218$ mGy, and $186$ mGy, respectively. The predicted whole-body dose for both female and male astronauts is $219$ mGy and $215$ mGy, respectively.

For an exploration mission to Mars, with reference parameters as described above including 400 days in transit and 500 days of surface exploration, the predicted whole body doses were found to be 503.3 mGy and 492.5 mGy for the female and male astronauts, respectively. For the female astronaut, detailed cumulative organ specific doses included 469.0 mGy predicted dose to the gastrointestinal system, 512.6 mGy to the circulatory system, and 476.2 mGy to the brain. For the male astronaut, the predicted cumulative doses for the same organ systems were 446.3 mGy, 501.8 mGy, and 426.7 mGy, respectively.


\section*{Discussion}

Whole-body daily absorbed dose as calculated by the computational phantom approach was within $6.7$\% of actual recorded ISS values \cite{berger2016dosis, berger2017dosis, kroupa2015semiconductor}. The remaining error in caculated values is most likely attributed to the spectrum input; here, calculations used measured particle flux and energy averaged over several days. This calculation may be improved in future iterations by refinement of computational inputs and variability derived from considerations such as the solar cycle, shielding, mission parameters and trajectory, and other factors that may alter particle fluence and intravehicular environmental complexity. Even so, these calculations demonstrate a high degree of accuracy particularly when compared to previous approaches.  Kuznetsov et. al. compared predicted to measured absorbed dose using a numerical model, SNIP-2016, to calculate the energy spectra of GCR heavy charged particles \cite{kuznetsov2017empirical}. They showed a 30\% deviation between calculated and measured absorbed dose values from dosimeters \cite{kuznetsov2019comparison}. Our results are a significant improvement on previous results.  

This work demonstrates the feasibility of performing accurate 3D Monte Carlo simulations to predict the dose topology following an exposure to complex space radiation fields on high-fidelity anatomical male and female tetrahedral-type computational phantoms. Using novel, high-fidelity phantoms with internal physiology accurate to the scale of a micron in conjunction with the moderator block method \cite{chancellor2017targeted}, this approach demonstrates use of computer-based and beamline-based space radiation calculations to predict on-orbit exposures with a high degree of accuracy. Our approach can thus be introduced into existing Monte Carlo codes and applied to various spaceflight mission scenarios.

While advanced surface phantoms have been used in clinical scenarios, their prior use has been limited to NURBS-based models in space radiation applications. Surface phantoms can be generated with relative ease from medical imaging (for example, computed tomography (CT) imaging), providing high resolution sources for phantom production and dosimetric calculations. As computational capacity expands in speed and accessibility, high-precision simulations utilizing mesh-phantoms can be performed to provide accurate assessments for highly specific radiation scenarios. The results of this work show the utility of using high performance computers and sophisticated computational models for assessing spaceflight-relevant scenarios. Areas of particular interest for radiobiology and effective dose calculations can be modeled, and dose deposition and track structure analysis on the order of microns and smaller can be simulated and investigated. Mesh computational phantoms have little limit on their resolution capacity, providing the groundwork for computational simulations investigating biological effects on increasingly smaller scales. Areas of interest for adverse radiation effects, such as mucosal linings, lymph nodes, glands, etc., can be modeled in this manner and provide accurate calculations specific to body tissue or organ system. 

As mentioned above, surface-type and tetrahedral-type phantoms can be generated from CT or other medical imaging data. As the complexity of structures and layer resolution exceeds imaging capacity, additional modeling input for the phantom would be needed; even so, whole organ systems and well-characterized subregions of organs can be segmented and transformed into a tetrahedral-type phantom. Medical imagery inputs into phantom design and dose-exposure calculations could improve upon personalization of risk assessments for future spaceflight. In essence, this overall capability could enable assessment of individualized exposures that would be incurred by astronaut crews during spaceflight and facilitate the application of a personalized risk assessment in order to better predict and communicate the short- and long-term health outcomes of space radiation exposures.

\section*{Conclusions}

This novel approach utilizing high-fidelity anatomical tetrahedral-type computational phantoms in conjunction with the unique moderator block beamline radiation calculation provides a method of predicting daily and cumulative space radiation dose-exposures to individual organs and whole body exposures with a remarkable degree of accuracy. Tailoring of this method to further improve upon accuracy offers the opportunity for rapid advancement in predictive capability of radiation exposure risks. This could potentially allow for vastly improved understanding of cumulative dose expectations, specific organ exposures and sequelae, and, ultimately, prediction of the impacts and risks associated with space radiation exposure specific to individual crewmembers and design reference missions. As human spaceflight expands outside of LEO, such predictive capability may revolutionize the understanding of health and performance risks and allow for advancement of preventive capabilities to limit health impacts from cumulative space radiation exposure.

\section*{Author contributions statement}
J.C.C. and M.C. conceived, conducted, and analyzed the experiment. R.S.B. and S.A.C.~contributed to the medical discussion and interpretations. All authors contributed to the development and review of the manuscript. 

\section{Additional information}
J.~C.~C. and M.~C have filed a U.S.Provisional Patent, No. 63/316,832, based on the methods described.

\section*{Acknowledgements}\footnotesize
\noindent Parts of this work were supported by the National Aeronautics and Space Administration through grants 80NSSC21K0544 and 80NSSC21K0273, and LaSPACE through grant number 80NSSC20M0110. The authors would like to thank the Center for Computation \& Technology (CCT) at Louisiana State University and the Texas Advanced Computing Center (TACC) at The University of Texas at Austin for providing high performance computing resources that have contributed to the research results reported within this paper.
Additionally, the authors would like to thank Hanyang University Radiation Engineering Laboratory (HUREL) at Hanyang University in Seoul, Korea whose phantoms were used in this work. 

\section*{Competing Interests}\footnotesize
\noindent The authors declare no competing financial or non-financial interests.

\section*{Data Availability}\footnotesize
\noindent The datasets generated and analyzed are available from the corresponding author on reasonable request.

\setlength{\arrayrulewidth}{0.1mm}
\setlength{\tabcolsep}{30pt}
\renewcommand{\arraystretch}{1.5}
\begin{table*}[t]
\begin{center}
\begin{tabular}{ |p{2.9cm}|p{2.7cm}|p{2.8cm}|}
\hline
Organ System & \multicolumn{2}{|c|}{Male Phantom} \\
\hline\hline
 & \emph{Absorbed \newline Dose (mGy/day)} & \emph{Energy \newline Deposited (mJ/day)}\\
 \hline\hline
\textbf{Whole Body Dose} & $\mathbf{0.2985\:\pm{\:0.0002}}$ & $\mathbf{32.42\:\pm{\:0.5}}$\\
 \hline \hline
\emph{Skeletal System} & 0.2778 $\pm{\:0.0002}$ & 3.533 $\pm{\:0.003}$\\\hline
Spine & 0.2722 $\pm{\:0.0004}$ & 0.576 $\pm{\:0.001}$\\
Humeri & 0.281 $\pm{\:0.001}$ & 0.1945 $\pm{\:0.0007}$\\
Femur & 0.2754 $\pm{\:0.0006}$ & 0.510 $\pm{\:0.001}$\\
\hline\hline
\emph{GI System} & 0.2705 $\pm{\:0.0003}$ & 1.867 $\pm{\:0.002}$\\\hline
Oral Cavity & 0.284 $\pm{\:0.003}$ & 0.0256 $\pm{\:0.0003}$ \\
Esophagus & 0.323 $\pm{\:0.002}$ & 0.0315 $\pm{\:0.0002}$\\
Stomach & 0.270 $\pm{\:0.001}$ & 0.1567 $\pm{\:0.0006}$ \\
Small Intestine & 0.2730 $\pm{\:0.0005}$ & 0.4328 $\pm{\:0.0008}$ \\
Colon & 0.2582 $\pm{\:0.0006}$ & 0.2676 $\pm{\:0.0006}$\\
\hline\hline
\emph{Circulatory System} & 0.3041 $\pm{\:0.0004}$ & 0.890 $\pm{\:0.001}$\\\hline
Heart & 0.2978 $\pm{\:0.0009}$ & 0.349 $\pm{\:0.001}$ \\
Blood Vessels & 0.3084 $\pm{\:0.0005}$ & 0.5407 $\pm{\:0.0008}$\\
\hline\hline
\emph{Lymphatic Nodes}& 0.293 $\pm{\:0.001}$ & 0.0720 $\pm{\:0.0002}$\\
\hline\hline
\emph{Urogenital System} & 0.2947 $\pm{\:0.0008}$ & 0.2933 $\pm{\:0.0008}$\\\hline
Kidney & 0.276 $\pm{\:0.001}$ & 0.1523 $\pm{\:0.0006}$\\
Uretha & 0.298 $\pm{\:0.002}$ & 0.00646 $\pm{\:0.00005}$ \\
Bladder & 0.312 $\pm{\:0.002}$ & 0.1023 $\pm{\:0.0005}$\\
Gonads & 0.388 $\pm{\:0.005}$ & 0.0187 $\pm{\:0.0002}$ \\
Prostate/Uterus & 0.316 $\pm{\:0.004}$ & 0.0072 $\pm{\:0.0001}$\\
\hline\hline
\emph{Respiratory System} & 0.3002 $\pm{\:0.0008}$ & 0.488 $\pm{\:0.001}$\\\hline
Trachea & 0.350 $\pm{\:0.005}$ & 0.4744 $\pm{\:0.0001}$\\
Lungs & 0.2999 $\pm{\:0.0008}$ & 0.468 $\pm{\:0.001}$\\
Bronchi & 0.328 $\pm{\:0.003}$ & 0.00137 $\pm{\:0.00002}$\\
\hline\hline
Breast & 0.303 $\pm{\:0.003}$ & 0.0139 $\pm{\:0.0001}$\\
Skin & 0.3220 $\pm{\:0.0004}$ & 1.400 $\pm{\:0.002}$\\
Eyes & 0.303 $\pm{\:0.006}$ & 0.0052 $\pm{\:0.0001}$\\
Brain & 0.259 $\pm{\:0.001}$ & 0.940 $\pm{\:0.002}$\\
Pituitary Gland & 0.25 $\pm{\:0.01}$ & 0.00017 $\pm{\:0.00009}$\\
Spinal Cord & 0.292 $\pm{\:0.002}$ & 0.0142 $\pm{\:0.0002}$ \\
Thymus & 0.348 $\pm{\:0.004}$ & 0.0181 $\pm{\:0.0003}$ \\
Thyroid & 0.371 $\pm{\:0.005}$ & 0.0134 $\pm{\:0.0003}$ \\
Muscle & 0.2992 $\pm{\:0.0003}$ & 11.89 $\pm{\:0.06}$\\
\hline
\end{tabular}
\end{center}
\caption{\textbf{Organ and organs system doses and energy deposition for the male phantom.} Whole organ doses were determined using volume weighted averages of the calculated regions while organ system doses are the sum of the calculated major organs in each system. The humeri, femur, and lung values are the average values of both right and left sides. The blood vessels value is an average of the doses to the veins and the arteries, and the value for the eyes is the average of the left and right eyes. Energy deposited was determined by using the value for absorbed dose and the mass of the individual regions.}
\label{tablemale}
\end{table*}

\begin{table*}[t]
\begin{center}
\begin{tabular}{ |p{2.9cm}|p{2.7cm}|p{2.8cm}|}
\hline
Organ System & \multicolumn{2}{|c|}{Female Phantom} \\
\hline
 & \emph{Absorbed \newline Dose ($m$Gy/day)} & \emph{Energy \newline Deposited ($m$J/day)}\\
 \hline
{\textbf{Whole Body Dose}} & $\mathbf{0.3050\:\pm{\:0.0002}}$ & $\mathbf{28.7\:\pm{\:0.5}}$\\
 \hline\hline 
\emph{Skeletal System} & 0.2873 $\pm{\:0.0003}$ & 2.710 $\pm{\:0.002}$\\\hline
Spine & 0.2850 $\pm{\:0.0005}$ & 0.4785 $\pm{\:0.0009}$\\
Humeri & 0.286 $\pm{\:0.001}$ & 0.1575 $\pm{\:0.0007}$\\
Femur & 0.2598 $\pm{\:0.007}$ & 0.3288 $\pm{\:0.0008}$\\
\hline\hline
\emph{GI System} & 0.2842 $\pm{\:0.0004}$ & 1.7 $\pm{\:0.4}$\\\hline
Oral Cavity & 0.337 $\pm{\:0.004}$ & 0.0246 $\pm{\:0.0003}$ \\
Esophagus & 0.322 $\pm{\:0.002}$ & 0.0273 $\pm{\:0.0001}$\\
Stomach & 0.274 $\pm{\:0.001}$ & 0.1461 $\pm{\:0.0006}$ \\
Small Intestine & 0.2863 $\pm{\:0.0006}$ & 0.3923 $\pm{\:0.0007}$ \\
Colon & 0.2717 $\pm{\:0.0006}$ & 0.2769 $\pm{\:0.0006}$\\
\hline\hline
\emph{Circulatory System} & 0.3107 $\pm{\:0.0005}$ & 0.673 $\pm{\:0.001}$\\\hline
Heart & 0.3017 $\pm{\:0.0009}$ & 0.2638 $\pm{\:0.0008}$ \\
Blood Vessels & 0.3167 $\pm{\:0.0006}$ & 0.4098 $\pm{\:0.0007}$\\
\hline\hline
\emph{Lymphatic Nodes} & 0.301 $\pm{\:0.001}$ & 0.0591 $\pm{\:0.0002}$\\
\hline
\emph{Urogenital System} & 0.2936 $\pm{\:0.0008}$& 0.2809 $\pm{\:0.0008}$\\
Kidney & 0.284 $\pm{\:0.001}$ & 0.1342 $\pm{\:0.0005}$\\
Uretha & 0.299 $\pm{\:0.002}$ & 0.00609 $\pm{\:0.00004}$ \\
Bladder & 0.302 $\pm{\:0.002}$ & 0.0961 $\pm{\:0.0005}$\\
Gonads & 0.293 $\pm{\:0.004}$ & 0.00490 $\pm{\:0.00007}$ \\
Prostate/Uterus & 0.309 $\pm{\:0.002}$ & 0.0335 $\pm{\:0.0003}$\\
\hline\hline
\emph{Respiratory System} & 0.3069 $\pm{\:0.0007}$ & 0.396 $\pm{\:0.001}$\\\hline
Trachea & 0.345 $\pm{\:0.004}$ & 0.00375 $\pm{\:0.00005}$\\
Lungs & 0.3063 $\pm{\:0.0008}$ & 0.385 $\pm{\:0.001}$\\
Bronchi & 0.327 $\pm{\:0.005}$ & 0.00058 $\pm{\:0.00009}$\\
\hline\hline
Breast & 0.2899 $\pm{\:0.0009}$ & 0.2579 $\pm{\:0.0008}$\\
Skin & 0.3260 $\pm{\:0.0004}$ & 1.018 $\pm{\:0.001}$\\
Eyes & 0.266 $\pm{\:0.006}$ & 0.00460 $\pm{\:0.0001}$\\
Brain & 0.289 $\pm{\:0.001}$ & 0.421 $\pm{\:0.002}$\\
Pituitary Gland & 0.31 $\pm{\:0.02}$ & 0.00022 $\pm{\:0.00001}$\\
Spinal Cord & 0.315 $\pm{\:0.003}$ & 0.00775 $\pm{\:0.00008}$ \\
Thymus & 0.327 $\pm{\:0.004}$ & 0.0089 $\pm{\:0.0001}$ \\
Thyroid & 0.357 $\pm{\:0.005}$ & 0.0092 $\pm{\:0.0001}$ \\
Muscle & 0.3012 $\pm{\:0.0003}$ & 7.070 $\pm{\:0.007}$\\
\hline
\hline
\end{tabular}
\end{center}
\caption{\textbf{Organ and organs system doses and energy deposition
for the female phantom.} Whole organ doses were determined using volume weighted averages of the calculated regions while organ system doses are the sum of the calculated major organs in each system. The humeri, femur, and lung values are the average values of both right and left sides. The blood vessels value is an average of the doses to the veins and the arteries, and the value for the eyes is the average of the left and right eyes. Energy deposited was determined by using the value for absorbed dose and the mass of the individual regions.}
\label{tablefemale}
\end{table*}
\end{document}